\documentstyle[12pt,fleqn]{article}
\input psfig
\def\be{\begin{equation}}
\def\ee{\end{equation}}
\def\bea{\begin{eqnarray}}
\def\eea{\end{eqnarray}}
\def\la{\mathrel{\mathpalette\fun <}}
\def\ga{\mathrel{\mathpalette\fun >}}
\def\fun#1#2{\lower3.6pt\vbox{\baselineskip0pt\lineskip.9pt
        \ialign{$\mathsurround=0pt#1\hfill##\hfil$\crcr#2\crcr\sim\crcr}}}
\def\re#1{{[\ref{#1}]}}
\def\ret#1#2{{[\ref{#1},\ref{#2}]}}

\def\thesection{}
\def\rhoeq{{{\rho_{\rm EQ}}}}
\def\rhodm{{{\rho_{\rm DM}}}}
\def\Teq{{{T_{\rm EQ}}}}
\def\aeq{{{a_{\rm EQ}}}}

\begin{document}
\thispagestyle{empty}
\def\pag{{{\langle P \rangle}}}
\null\vspace{-42pt}
\begin{flushright}
\baselineskip=12pt
{\footnotesize
FERMILAB--PUB--94/055--A\\
astro-ph/yymmxxx\\
March, 1994}
\end{flushright}
\renewcommand{\thefootnote}{\fnsymbol{footnote}}
\baselineskip=24pt

\vspace{42pt}
\begin{center}
{\Large \bf Large-amplitude isothermal fluctuations \\
		and high-density dark-matter clumps}\\
\vspace{1.0cm}
\baselineskip=14pt

Edward W.\ Kolb\footnote{Electronic mail: {\tt rocky@fnas01.fnal.gov}}\\
{\em NASA/Fermilab Astrophysics Center, \\
Fermi National Accelerator Laboratory, Batavia, IL~~60510, and\\
Department of Astronomy and Astrophysics, Enrico Fermi Institute\\
The University of Chicago, Chicago, IL~~ 60637}\\
\vspace{0.4cm}
Igor I.\ Tkachev\footnote{Electronic mail: {\tt tkachev@fnas13.fnal.gov}}\\
{\em NASA/Fermilab Astrophysics Center\\
Fermi National Accelerator Laboratory, Batavia, IL~~60510, and\\
Institute for Nuclear Research of the Academy of Sciences of
Russia\\Moscow 117312, Russia}\\
\end{center}

\baselineskip=24pt

\begin{quote}
\hspace*{2em} Large-amplitude isothermal fluctuations in the dark matter energy
density, parameterized by $\Phi\equiv\delta\rhodm/\rhodm$, are studied within
the framework of a spherical collapse model. For $\Phi \ga 1$, a fluctuation
collapses in the radiation-dominated epoch and produces a dense dark-matter
object. The final density of the virialized object is found to be $\rho_F
\approx 140\, \Phi^3 (\Phi+1) \rhoeq$, where $\rhoeq$ is the matter density at
equal matter and radiation energy density.  This expression is valid for the
entire range of possible values of $\Phi$, both for $\Phi \gg 1$ and $\Phi \ll
1$. Some astrophysical consequences of high-density dark-matter clumps are
discussed.
\vspace*{12pt}

PACS number(s): 98.80.Cq, 14.80.Gt, 05.30.Jp, 98.70.-f

\end{quote}
\newpage
\baselineskip=24pt
\setcounter{page}{2}
\renewcommand{\thefootnote}{\arabic{footnote}}
\addtocounter{footnote}{-2}

\thesection{\centerline{\bf I. INTRODUCTION}}
\setcounter{section}{1}
\setcounter{equation}{0}
\vspace{18pt}

In almost all modern cosmological models, galaxies, clusters, and all
large-scale structures develop through the gravitational instability of
small-amplitude, seed density fluctuations.   In most of these models cold dark
matter is an important constituent of the total mass density of the Universe.
There are two basic types of seed density fluctuations, curvature and
isocurvature,\footnote{The division of fluctuations into curvature and
isocurvature is strictly true only outside the Hubble radius \re{myths}.} and
in general, both are expected to be produced in the early Universe.  By
definition, the total energy density in an isocurvature fluctuation is
constant; the fluctuation is in the relative contribution to the total energy
density of different components in a multicomponent system.  Important examples
of this type are the fluctuations induced in the baryons by some dissipative
process in a Universe containing both baryons and dark matter, and topological
or non-topological field configurations like cosmic strings or textures. While
the amplitude of either type of fluctuation on large scales is strongly
restricted by microwave background anisotropy constraints, the amplitude of
small-scale fluctuations can be large, even non-linear, at the epoch of last
scattering.   The spectrum of small-scale fluctuations do not necessarily have
to reflect the shape of the power spectrum of the primordial fluctuations
generated at the inflationary epoch, since the small-scale fluctuations may
well be generated later, e.g., during various cosmological phase transitions.

In this paper we are interested in isocurvature fluctuations that enter the
horizon before the temperature of equal energy densities of matter and
radiation, $\Teq = 5.5 \,\Omega_0 h^2 \, {\rm eV}$ \re{kt90}.  We will consider
scales much smaller than the horizon, so the radiation energy density should be
homogeneous.

It is well known \re{pm74} that the growth of small-amplitude isothermal
fluctuations is suppressed by the cosmological expansion, and the fluctuations
do not grow until after the equality epoch. However, this is true only in
linear theory.  The self gravity of large-amplitude, non-linear fluctuations
may become important before $\Teq$, and consequently they collapse earlier.
Therefore they are capable of producing very dense objects after they separate
out from the general expansion and virialize.

We refer to these isothermal fluctuations as ``clumps.''\footnote{Since the
clumps may be very dense compared to the background, we do not refer to them as
``perturbations.''}   Let us specify the density of a dark-matter clump as
\begin{equation}
\delta \rhodm/\rhodm = \Phi,
\label{defphi}
\end{equation}
where $\Phi$ is not necessarily small. For example, ``typical'' axion
miniclusters \re{hr88} will have $\Phi \sim 1$. In  Ref.\ \re{kt93} it was
found that accounting for non-linear effects in the evolution of axions at the
crucial epoch when the axion mass switches on can lead to considerably larger
density in many miniclusters, with $\Phi$ in the range $1$ to $10^4$.
Dark-matter clumps seeded by wakes induced by cosmic strings or by textures
also will have $\Phi \sim 1$ \re{ss93}. {\em Seeded clumps are particularly
interesting in the case of WIMP dark matter }

It was pointed out in Ref.\ \re{kt93} that the final virialized density in a
clump has to scale as $\rho_F \sim \Phi^4 \rhoeq$. Because of the dependence
upon the fourth power of $\Phi$, even a small increase in $\Phi$ is very
important.  For the same reason, the final density can be sensitive to the
details of the evolution of the clump in the radiation-dominated era. To our
knowledge, a detailed study of the non-linear evolution of large-amplitude
isothermal fluctuations has never been performed. However, it is very important
in various phenomenological implications including both direct and non-direct
dark matter searches.  In this paper we consider this problem.

The clumpiness of the dark matter has important implications for attempts to
detect dark matter.  Clearly the signal in direct detection experiments for
dark matter is proportional to the dark matter density.   For the rare direct
encounter with a clump,  there could be a huge amplification of the signal.
However, if the clumpiness is too high, the flux of unclumped dark matter will
be too small for a reasonable detection rate. The rate of WIMP annihilation
contributing to the $\gamma$-ray background \ret{sb87}{is87} is proportional to
the density as well.  In the case of clumped dark matter, there will be
stronger constraints on dark matter from indirect searches. In very dense axion
clumps, Bose star formation becomes possible \re{it91} (clumps with $\Phi \sim
30$ already satisfy the critical condition for this \re{kt93}), which in turn
can lead to the formation of radio sources \re{it86}. Another possible
manifestation of high density clumps is the phenomenon of microlensing.

To study the structure and evolution of high density clumps, a full
3-dimensional numerical simulation is needed. However, for an isolated clump
some relevant physical information can be extracted from a 1-dimensional
spherical model. The spherical model proved useful in studies of the
gravitational non-linear evolution in the epoch of matter domination when it is
possible to find exact analytic solutions \re{p80}. In the present paper we
generalize this model to include radiation. Although there are no analytic
solutions, the result turns out to be very simple: The final density in a
virialized clump is $\rho_F \approx 140\, \Phi^3 (\Phi+1) \rhoeq$ in the whole
range of possible values of $\Phi$, both for $\Phi \gg 1$ and for $\Phi \ll 1$.

\vspace{48pt}
\thesection{\centerline{\bf II. A SPHERICAL MODEL}}
\setcounter{section}{2}
\setcounter{equation}{0}
\vspace{18pt}

Let us consider a spherical region of radius $r$ containing an overdensity of
pressureless matter in an expanding Universe.  In a spatially flat Universe,
every overdense region eventually reaches some maximum size and re-collapses.
The total mass of matter in the region inside $r$, $M_{\rm TOT}$, is an
integral of the motion so long as the region expands. Since we will consider
scales  much smaller than the Hubble radius, we can consider the radiation to
be homogeneous, with its time evolution determined by the general expansion of
the Universe, and not by the local conditions.

The equation of motion for the radius of the region is
\begin{equation}
\ddot{r}=-\frac{8\pi G}{3} \rho_R  - {G M_{\rm TOT} \over r^2} \, .
\label{sh1}
\end{equation}
It is convenient to change to the conformal time coordinate, $d\eta = dt/a(t)$,
and then to rewrite this equation of motion in the comoving reference frame, $r
= a(\eta) R_\xi (\eta) \xi$, where $\xi$ is the comoving label of a given shell
and $R_\xi (\eta)$ measures the deviation of the shell motion from the uniform
Hubble flow of the background Friedmann Universe. In what follows we shall omit
the subscript $\xi$ on $R(\eta)$, but it should be understood that there is a
separate evolution for each shell.

We shall assume that the scale factor $a(\eta)$ satisfies the Einstein
equations for an $\Omega_0 =1$ Universe filled with radiation and pressureless
matter:
\begin{equation}
a^{\prime 2} = {8 \pi G \over 3} (\rho_M + \rho_R) a^4\ ; \qquad
a^{\prime \prime} = {4\pi G \over 3}\rho_M a^3 \, ,
\label{eeq}
\end{equation}
where prime denotes $d/d\eta$. We parametrize the radiation and matter energy
densities as, $\rho_R=\rhoeq(\aeq/a)^4$ and $\rho_M=\rhoeq(\aeq/a)^3$.
The solution to the background equations, Eqs.\ (\ref{eeq}), is
\begin{equation}
a(\eta)=\aeq[2(\eta /\eta_*)+(\eta /\eta_*)^2] \, ,
\label{scf}
\end{equation}
where $\eta_*^{-2}=2\pi G \rhoeq \aeq^2/3$.

The equation of motion [Eq.\ (\ref{sh1})] in terms of conformal time is
\begin{equation}
aR^{\prime\prime} +a^\prime R^\prime +\left({GM_{\rm TOT}\over \xi^3 R^2}
-{4\pi G\over 3} a^3\rho_M R\right)=0 \, .
\label{sh2}
\end{equation}
The radiation energy density does not enter this equation explicitly, but its
effect is encoded in the evolution of the scale factor.  We also parametrize
the total mass of matter inside the shell in terms of the excess over the
homogeneous background, denoted as $\Phi(\xi) \equiv \delta\rho_M/\rho_M$.  The
total mass within the region is
\begin{equation}
M_{\rm TOT} \equiv {4\pi \over 3} \rhoeq \aeq^3[1+\Phi (\xi )]\xi^3 \, .
\label{mtot}
\end{equation}
Changing from $\eta$ to $x \equiv a/\aeq$ as the independent variable, we
finally obtain
\begin{equation}
x(1+x)\frac{d^2R}{dx^2}+\left(1+\frac{3}{2}x\right)\frac{dR}{dx} +\frac{1}{2}
\left(\frac{1+\Phi}{R^2} -R\right)=0 \, .
\label{shf}
\end{equation}

This equation reduces to the Meszaros equation \re{pm74} in the limit of small
deviation of the shell motion from the general cosmological expansion, $R
\equiv 1-\delta$ and $\delta \ll 1$, if we assume no excess in total mass of
the matter, i.e., $\Phi =0$,
\begin{equation}
x(1+x)\frac{d^2\delta}{dx^2} +\left(1+\frac{3}{2}x\right)\frac{d\delta}{dx}
-\frac{3}{2}\delta =0 \,  .
\label{meq}
\end{equation}
The latter is hypergeometric equation, and its growing mode is $\delta
=\delta_0 (1+3x/2)$, implying the well known result that the growth of small
fluctuations is significant only after the Universe becomes matter dominated.

We have solved Eq.\ (\ref{shf}) numerically, assuming $R(x_0)=1$ at some early
time, $x_0 \ll 1$. Note that at small $x$, the second derivative in Eq.\
(\ref{shf}) can be neglected, and the solution with initial conditions fixed at
$x_0 =0$ is $R = (1-3\Phi x/2 )^{1/3} \simeq 1-\Phi x/2 $, where the expansion
is justified since the solution is valid only at small $x$. Actually, this is
the separatrix, i.e., independently of the initial value of $R^\prime$, all
solutions tend to it (provided $x_0 \ll 1$). The results of numerical
integration of Eq.\ (\ref{shf}) proved to be insensitive to $R^\prime (x_0)$
already at $x_0\Phi < 10^{-3}$.  For several different values of $\Phi$ they
are shown in Fig.\ 1.

It is possible to find an analytic approximation to $R(x)$ as a power series in
$x$ to any given order. To third order it is
\begin{equation}
R = 1-\frac{\Phi x}{2}-\frac{\Phi^2 x^2}{8}-\frac{(8\Phi^3 -\Phi^2)x^3}{144} \,
{}.
\label{sp}
\end{equation}
The first three terms in this decomposition provide a good practical fit to the
solution. This fit is shown in Fig.\ 1 by the dotted line. The last term in
Eq.\ (\ref{sp}) shows that the solution is not simply a function of the product
$\Phi x$.

Our main goal is to find the parameters of the fluctuation, i.e., its radius
and density, at the moment when the fluctuation turns around. For later times
the assumption of spherical symmetry breaks down; however, we can assume that
the radius of a virialized gravitationally bound object will be one-half of the
turnaround radius, and the density inside the object will be eight times larger
than the density at turnaround \re{p80}. The turnaround time, defined by
$\dot{r}=0$, in coordinates of Eq.\ (\ref{shf}) is the solution to the equation
$R +x\, dR/dx =0$. Up to order ${\cal O}(x^2)$ the function $R$ only depends
upon the product $\Phi x$, so to this order in $x$ the scale factor at
turnaround will be given by $x_{\rm TA} = {\rm const}/\Phi$ and to the same
order in $x$, $R_{\rm TA}={\rm const}$. The matter density of a fluctuation at
turnaround is $\rho_{\rm TA}=(1/4\pi r^2)dM/dr$, with $r= R_{\rm TA}x_{\rm
TA}\xi$, so  it is appropriate to represent the parameters at turnaround as
\begin{equation}
x_{\rm TA} = C_x/\Phi  \, ; \qquad
\rho_{\rm TA} = C_\rho \rhoeq \frac{\Phi^3}{3\xi^2} \frac{d}{d\xi}(1+\Phi
)\xi^3  \, \, .
\label{nr}
\end{equation}
We expect only a weak dependence of $C_x$ and $C_\rho$ upon $\Phi$, since
higher order corrections are small. Results of a numerical integration for
$C_x$ and $C_\rho$ are shown in Fig.\ 2 and Fig.\ 3 respectively. These figures
demonstrate that for practical applications we can consider $C_x$ and $C_\rho$
to be constants in entire range of possible values of $\Phi$, both $\Phi \gg 1$
and $\Phi \ll 1$.

Let us compare our results to predictions of a standard spherical model, which
is valid for the matter-dominated epoch, i.e., at small $\Phi$. The spherical
model predicts $\rho_{\rm TA}/\rho_{\rm B}= 9\pi^2/16$ for the density contrast
at turnaround, where $\rho_{\rm B}$ is the background density \re{p80}. Using
Fig.\ 2 we can extract the corresponding values of $\rho_{\rm B}$ and then
calculate the standard spherical model prediction for $C_\rho$. This is
presented in Fig.\ 3 by the dashed line. Both dashed and solid lines coincide
in the limit of small $\Phi$, as they must.

\vspace{48pt}
\thesection{\centerline{\bf III. APPLICATIONS}}
\setcounter{section}{3}
\setcounter{equation}{0}
\vspace{18pt}

With the function $\Phi (\xi )$ in the general form we have assumed, the
effective radius after virialization for each shell of a given label $\xi$ will
be half the turnaround radius of the shell.\footnote{Strictly speaking this is
not always true, but for the practical applications we will consider below the
approximation should be reasonable.}  Using Eq.\ (\ref{nr}) we obtain for the
final density profile in a virialized object
\begin{equation}
\rho_F \approx 140 \rhoeq \frac{\Phi^3}{3\xi^2} \frac{d}{d\xi}(1+\Phi )\xi^3,
\label{dp}
\end{equation}
where we have set $C_\rho \sim 17$.
For the core density, this formula gives $\rho_F \approx 140\, \Phi_0^3
(\Phi_0+1) \rhoeq$, where $\Phi_0 = \Phi (0)$. The numerical value of the
density at equality is $\rhoeq \approx 3 \times 10^{-16}\, (\Omega h^2)^4$g
cm$^{-3}$. Now let us turn to a few specific examples.

\vspace{16pt}
\centerline{{\bf A. Axion miniclusters}}

Fluctuations in the density of axions can be very high, possibly spanning the
range $1 \la \Phi \la 10^4$ \ret{kt93}{kt932}. Even with $\Phi$ as small as 1,
the density in miniclusters which form out of these fluctuations can be as much
as $10^{10}$ times larger than the local galactic halo density \re{hr88}.
Using Eq.\ (\ref{dp}), we obtain $\rho_F \sim 9 \times 10^{-14}\, \Omega^4 h^8
\, $g cm$^{-3}$ for $\Phi=1$. With $\Phi \gg 1$ this result must be multiplied
by $\Phi^4/2$.

The typical mass of an axion minicluster corresponds to the total mass in
axions within the horizon at $T \sim 1\, {\rm GeV}$ when the inverse axion mass
is equal to the Hubble length:  $M_{\rm MC} \sim 10^{-9}M_\odot$. The present
probability of a direct encounter with a minicluster is small, the encounter
rate is  1 per 30 million years with $\Phi=1$. Although the signal in an axion
detector \re{ps83} from a close encounter would be enormous, it might be a long
wait with a weak signal between encounters if a major fraction of the axions
are part of miniclusters.

There should be some miniclusters with $\Phi$ in the range $10^{-3}\la \Phi \la
1$. These collapse during the matter-dominated epoch and have a larger radii
than those with $\Phi\ga 1$ which collapsed in the radiation-dominared epoch,
so the probability of an encounter with a clump with $\Phi\ll1$ can be larger.
{}From the point of view of direct searches, even miniclusters with density
contrast of order two times the average with respect to the galactic halo
density are important.  Such miniclusters form just prior to the moment of
galaxy formation and started with $\Phi\sim10^{-3}$. For $\Phi<1$ the expected
time between encounters is given in terms of the number density of clumps, $n$,
the geometric cross section of the clump $\sigma\sim R_{\rm CLUMP}^2$, and the
virial velocity $v$ as
\begin{equation}
\tau = \frac{1}{n\sigma v}\simeq \frac{1}{\Phi}
\frac{\rho_F}{\rho_H}\frac{R_{\rm CLUMP}}{v},
\end{equation}
where $\rho_H$ is the halo density, and  $R_{\rm CLUMP}/v$ is the time the
Earth spends inside the minicluster.  The factor of $\Phi^{-1}$ appears because
the number density of miniclusters with $\Phi\ll1$ is suppressed in our model.
A minicluster with $\Phi\ll1$ would require the initial misalignment angle
$\theta$ (which is uniformally distributed  in the range $0$ to $2\pi$) to be
finely tuned to the mean misalignment angle to an accuracy
$\delta\theta/\theta\simeq \Phi/2$.  Using Eq.\ (\ref{nr}), we finally obtain
\begin{equation}
\tau = \Phi \tau_{\Phi=1} = \Phi \cdot 3\times10^{7}\mbox{yr}.
\end{equation}

Note that the miniclusters discussed so far appear if the axion field is
uncorrelated on scales larger than the Hubble radius at $T\sim 1$ GeV.  However
miniclusters with $\Phi\ll1$ can appear from primordial density fluctuations
generated by inflation without the suppression factor of $\Phi^{-1}$.  If this
is the case, then $\tau\simeq\Phi^2 \cdot 3\times10^7$ yr.  Since $\Phi$ would
be small, this would give a reasonable encounter rate, and the question of
formation and survival of small-scale clumps within the galaxy is worth further
study.

Another astrophysical outcome of very dense axion clumps can be the possibility
of  ``Bose star'' formation in axion miniclusters.  The Bose-Einstein
relaxation time in the minicluster due to axion self-interaction  is smaller
than the present age of the Universe with $\Phi \ga 30$ \ret{kt93}{it91}.

\vspace{16pt}
\centerline{{\bf B. Accretion by a point mass}}

The density profile in the halo accreted by a previously formed clump can be
calculated in the approximation of secondary infall onto an excess point mass
of mass $m$. In this case $\Phi (\xi )= m/M$, where $M=4\pi \rhoeq \aeq^3
\xi^3$ is the mass of the background dark matter within the shell with the
label $\xi$. Substituting this into Eq.\ (\ref{dp}) we find
\begin{equation}
\rho_F \approx 140 \rhoeq(m/M)^3 \, .
\label{dppm}
\end{equation}
This can be translated into $\rho_F$ as a function of $r$ since $M$ has to be
understood as the mass of dark matter residing within $r$. The result is
$\rho_F \propto r^{-9/4}$, the same power law one usually obtains for secondary
infall in the matter-dominated era \re{b}.  However, Eq.\ (\ref{dppm}) is valid
regardless of the time when collapse actually occurs.

\vspace{16pt}
\centerline{{\bf C. Cosmic strings}}

We can apply Eq.\ (\ref{dppm}) for clumps of dark matter seeded by loops of
cosmic strings so long as the peculiar velocity of a loop is sufficiently
small. Since the string loop is not a point object, this formula breaks down in
the region of small $M$. Namely, when the given shell turns around at $x_{\rm
TA} \approx 0.7M/m$, the loop size $l_S$ has to be smaller than the physical
radius of the shell, $r_{\rm TA} = x_{\rm TA}R_{\rm TA} \xi \aeq$, for Eq.\
(\ref{dppm}) to be valid. This gives the restriction $M \ga 3 m^{3/2}
\mu^{-3/4} \rhoeq^{1/4} \equiv M_C$, where  $m= \mu l_S$. We can consider $M_C$
as the mass of the core region. The corresponding maximum density which can be
achieved in the core is $\rho_C \sim 15\sqrt{\mu^3\rhoeq}/M_C$. This value of
the core density could be many orders of magnitude larger than the density at
equality. However, as we see from Eq.\ (\ref{dppm}), $\rho_F$ is much greater
than $\rhoeq$ only in the case when the mass of the string loop is larger than
the mass of the accreted dark matter. Consequently, with gradual loop decay due
to emission of gravitational radiation, the dark matter clump will
adiabatically expand and diminish in density.  This process of clump expansion
will continue untill $m \la M$. Since in the gravitational field the product
$rm$ is an adiabatic invariant for each dark  matter particle, where $r$ is the
effective radius of the orbit, we conclude that  in any clump for which $M < m$
initially, the present density will have the same order of magnitude,  $\rho
\sim 10^2 \rhoeq$. Dark-matter clumps seeded by wakes induced by long segments
of moving cosmic strings or by textures also will have $\Phi \sim 1$ (see
\re{ss93}), and correspondingly the same virialized density,  $\rho \sim 10^2
\rhoeq$. While this density is sufficiently high to be significant in
applications like annihilation of dark matter particles, it is too small to
cause microlensing, as we show below.

Cold dark matter accretion onto string loops both in the matter and radiation
dominated era was also considered in Ref.\ \re{as86}, however, only in the
linearized limit of Eq.\ (\ref{shf}), i.e., Eq.\ (\ref{meq}).

\vspace{16pt}
\centerline{{\bf D. WIMP annihilation}}

In the case where the dark-matter particle species is a  stable weakly
interacting massive particle (WIMP) such as a very massive neutrino or a
supersymmetric particle (photino, higgsino, or scalar neutrino), the WIMPs can
annihilate, contributing to the $\gamma$-ray flux. This places severe
constraints on the dark-matter density near the center of the galaxy
\ret{sb87}{is87}. Clumped dark-matter annihilation is even more efficient, and
places a very strong limit on the clumpines as a function of the WIMP
properties \re{ss93}.

For the $\gamma$-ray flux on Earth from WIMP annihilation in the clump, we can
write
\begin{equation}
I_\gamma \approx {\langle\sigma v\rangle \rho M \over 4 \pi r^2_\odot m^2_X} \,
,
\label{fl}
\end{equation}
where $r_\odot$ is the distance from the Earth to the clump ($r_\odot \approx
8.5 $ kpc is the distance to the center of the Galaxy), $m_X$ is the particle
mass, and $M$ is the mass of the clump. Since the particles are
non-relativistic, both in the clumps and at the epoch of cosmological
freeze-out of the WIMPs, the thermal average of the cross-section in the Eq.\
(\ref{fl}) is directly related to the cosmological abundance \re{lw77}:
\begin{equation}
\langle\sigma v\rangle \approx {4\times 10^{-27} \over \Omega_X h^2}\, {\rm
cm^3 sec^{-1}} \, .
\label{ca}
\end{equation}

If we consider a large region of (possibly) clumpy dark matter, like the
galactic core or spheroid, we must sum up the fluxes from each individual
clump. As a result, instead of $M$ in Eq.\ (\ref{fl}) we have to substitute
$\xi M_{\rm tot}$, where $\xi$ is the mass fraction of all clumps to the total
mass $M_{\rm tot}$ in the region.  For example, using $\rho \sim 10^2 \rhoeq$
we obtain for the central spheroid ($M_{\rm tot} \approx 10^8 M_\odot$)
\begin{equation}
I_\gamma \approx \xi \Omega_X^3 h^6 m_{20}^{-2}\, {\rm cm^{-2} sec^{-1}}\, ,
\label{fln}
\end{equation}
where $m_{20} \equiv m_X/20\, $ GeV. This has to be compared to the observed
upper limit to the $\gamma$-ray flux in the direction of the Galactic center
\re{sb87}, $I_\gamma \approx 4\times 10^{-7} {\rm cm^{-2} sec^{-1}}$.

\vspace{16pt}
\centerline{{\bf E. Gravitational microlensing}}

Two conditions must be satisfied for the clump to cause gravitational
microlensing \re{bp86}. First, the mass of the clump has to be in a range near
$0.1 M_\odot$. Second, the physical radius of the clump has to be smaller than
the Einstein ring radius, $R_{\rm E} =2\sqrt{GMd}$ where $d$ is the effective
distance to the lens (typically $d \sim 20 \, {\rm kpc}$). The second condition
restricts the density of the minicluster to be $\rho \ga 10^7
\rhoeq/\sqrt{M_{-1}}$, where $M_{-1} \equiv M/0.1 M_\odot$. If the lensing
object is a clump of non-interacting cold dark matter, it has to be formed from
a density fluctuation with $\Phi \ga 20$.

Dark matter clumps seeded by string loops or textures, which were considered in
Ref.\ \re{ss93}, are in the appropriate mass range; however, they have $\Phi
\sim 1$. Axion miniclusters can have $\Phi \ga 20$; however, they are too
light. While it is possible to invent models where both conditions are met for
some of the clumps (one example could be an axion model with an extremely
small, but non-zero, value for the $u$-quark mass), it is hardly likely that a
substantial amount of the dark matter has evolved into clumps capable of
lensing. On the other hand, anticipating significant numbers of microlensing
events (for the first positive reports see Ref.\ \re{macho}) in the future, it
is not excluded that some of them could be caused by the clumps in such classes
of models (especially if collisional relaxation is significant). The
corresponding light curve will be different from the MACHO event since clumps
are extended objects.

\vspace{12pt}

When our paper was almost completed we became aware of the paper Ref.\
\re{ps93} where Eq.\ (\ref{shf}) was studied in details; however, again for the
case which corresponds to $\Phi \ll 1$.

\vspace{12pt}

It is a pleasure to thank S. Colombi, A.\ Stebbins and  R.\ Caldwell for useful
discussions. This work was  supported in part by the DOE and NASA grant
NAGW--2381 at Fermilab.

\begin{picture}(400,50)(0,0)
\put (50,0){\line(350,0){300}}
\end{picture}

\vspace{0.25in}

\def\labelenumi{[\theenumi]}
\frenchspacing
\def\prl{{{\em Phys. Rev. Lett.\ }}}
\def\prd{{{\em Phys. Rev. D\ }}}
\def\pl{{{\em Phys. Lett.\ }}}
\begin{enumerate}

\item \label{myths} W. H. Press and E. T. Vishniac, {\em Ap. J.} {\bf 239}, 1
(1980).

\item \label{kt90} E. W. Kolb and M. S. Turner, {\em The Early Universe},
(Addison-Wesley, Redwood City, Ca., 1990).

\item \label{pm74}  P. Meszaros, Astron. Astrophys. {\bf 37}, 225 (1974).

\item \label{hr88} C. J. Hogan and M. J. Rees, Phys. Lett. {\bf B205}, 228
(1988).

\item \label{ss93} J. Silk and A. Stebbins, Astrophys. J. {\bf 411}, 439
(1993).

\item \label{kt93} E. Kolb and I. I. Tkachev, Phys. Rev. Lett. {\bf 71}, 3051
(1993).

\item \label{sb87} J. Silk and H. Bloemen, Astrophys. J. Lett. {\bf 313}, L47
(1987).

\item \label{is87} J. R. Ipser and P. Sikivie, Phys. Rev. D {\bf 35}, 3695
(1987).

\item \label{it91} I. I. Tkachev, Phys. Lett. {\bf B261}, 289 (1991).

\item \label{it86} I. I. Tkachev, Sov. Astron. Lett. {\bf 12}, 305 (1986);
I. I. Tkachev, Phys. Lett. {\bf B191}, 41 (1987).

\item \label{p80} P. J. E. Peebles, {\it The Large-Scale Structure of the
Universe} (Princeton Univ. Press, 1980).

\item \label{kt932} E. Kolb and I. I. Tkachev, {\it Non-liner axion dynamics
and formation of cosmological pseudo-solitons}, preprint FERMILAB-PUB-93/355-A
(unpublished).

\item \label{ps83} P. Sikivie, Phys. Rev. Lett. {\bf 51}, 1415 (1983).

\item \label{b} J. E. Gunn, Apstrophys. J. {\bf 218} (1977) 592.

\item \label{as86}
A. Stebbins, Astrophys. J. Lett. {\bf 303}, L21 (1986).

\item \label{lw77}
B. Lee and S. Weinberg, Phys. Rev. Lett. {\bf 39}, 165 (1977).

\item \label{bp86} B. Paczynski, Astrophys. J. {\bf 304}, 1 (1986); K. Griest,
Astrophys. J. {\bf 366} , 412 (1991).

\item \label{macho} E. Aubourg et al. (EROS Collaboration), {\it Nature}
(1993); C. Alcock et al. (MACHO Collaboration), {\it Nature} (1993).

\item \label{ps93} T. Padmanaban and K. Subramanian, Astrophys. J. {\bf 417}, 3
(1993).

\end{enumerate}
\newpage

\centerline{\bf FIGURE CAPTIONS:}
\vspace{36pt}

 Fig.\ 1:  Numerical solutions to Eq.\ (\ref{shf}) for several different values
of $\Phi$. The second-order fit is shown by the dotted line.

\vspace{24pt}
 Fig.\ 2:  The coefficient $C_x$ in Eq.\ (\ref{nr}) as a function of $\Phi$.

\vspace{24pt}
 Fig.\ 3:  The coefficient $C_\rho$ in Eq.\ (\ref{nr}) as a function of $\Phi$.
The dashed line is the prediction of the standard spherical model (matter
without radiation).

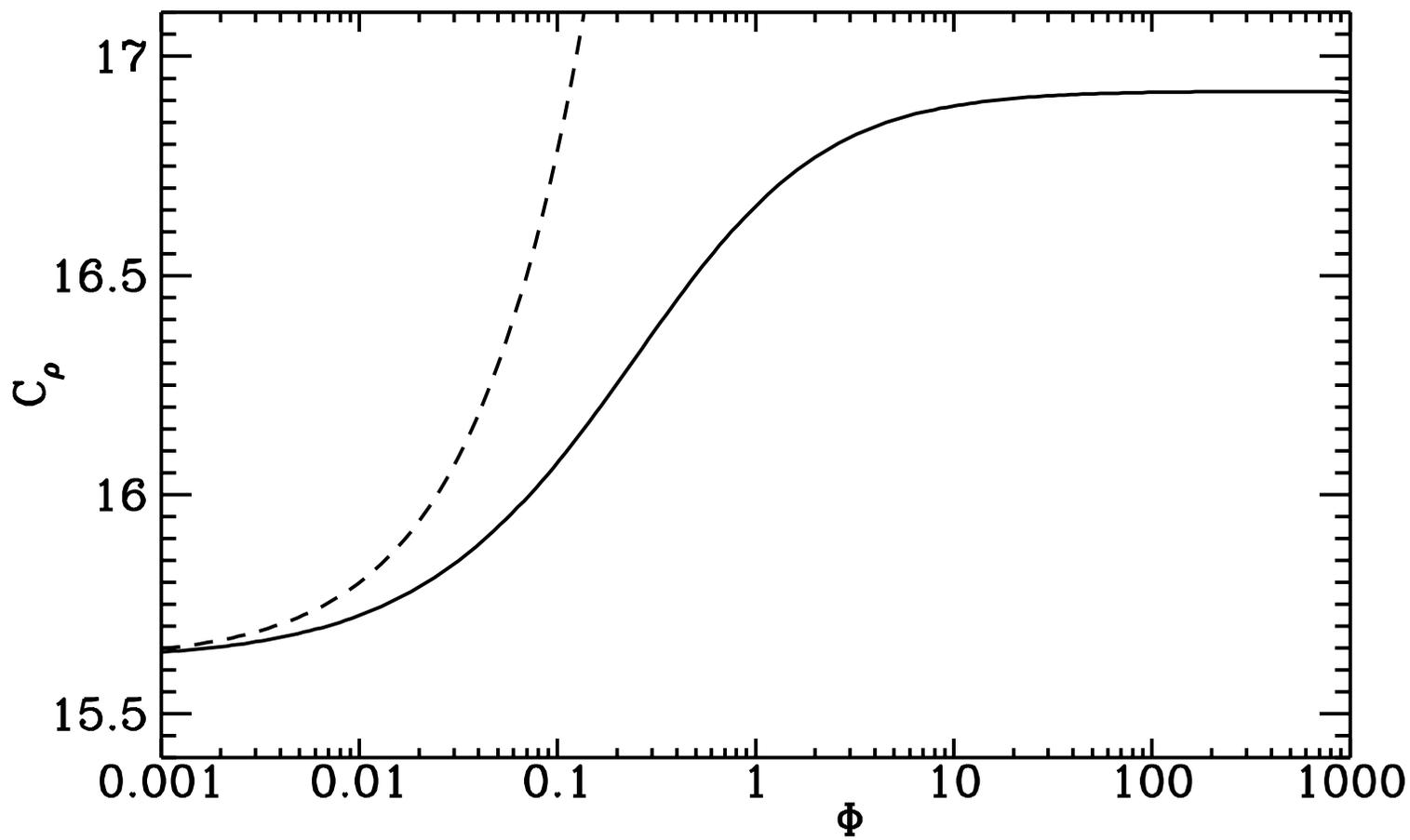
\end{document}